\documentclass[apj]{emulateapj}
\usepackage{amsmath}
\usepackage{natbib}
\usepackage{graphicx}

\newcommand{\beq}{\begin{equation}}
\newcommand{\eeq}{\end{equation}}
\newcommand{\bea}{\begin{array}}
\newcommand{\eea}{\end{array}}

\begin{document}

\title{Transits of Planets with Small Intervals in Circumbinary Systems}
\author{Hui-Gen Liu$^*$,Ying Wang, Hui Zhang and Ji-Lin Zhou}
\affil{School of Astronomy and Space Science and  Key Laboratory of Modern Astronomy\\ and Astrophysics in Ministry of Education, Nanjing
University, Nanjing, 210093, China; \\
huigen@nju.edu.cn}

\begin{abstract}
Transit times around single stars can be described well by a linear ephemeris. However, transit times in circumbinary systems are influenced both by the gravitational perturbations and the orbital phase variations of the central binary star. Adopting a coplanar analog of Kepler-16 as an example, we find that circumbinary planets can transit the same star more than once during a single planetary orbit, a phenomenon we call "tight transits". In certain geometric, the projected orbital velocity of the planet and the secondary star can approach zero and change sign, resulting in very long transits and/or 2-3 transits during a single binary orbit. Whether tight transits are possible for a particular system depends primarily on the binary mass ratio and the orbital architecture of both the binary and the planet. We derive a time-dependent criterion to judge when tight transits are possible for any circumbinary system. These results are verified with full dynamical integrations that also reveal other tight transit characteristics, i.e., the transit durations and the intervals between tight transits. For the seven currently known circumbinary systems, we estimate these critical parameters both analytically and numerically. Due to the mutual inclination between the planet and the binary, tight transits can only occur across the less massive star B in Kepler-16, -34, -35, and -47 (for both planets). The long-term average frequency of tight transits (compared to typical transits) for Kepler-16, -34, and -35 are estimated to be several percent. Using full numerical integrations, the next tight transit for each system is predicted and the soonest example appears to be Kepler-47b and -47c, which are likely to have tight transits before 2025. These unique and valuable events often deserve special observational scrutiny.
\end{abstract}

\section{Introduction}

Transits of exoplanets provide us a new way to detect exoplanets and to obtain the ratio of the transiting planetary radius to the occulted stellar radius.  As is well known, transit center times for planets around single stars can be fitted well linearly, and the small residuals of linear fitting are known as transit time variations (hereafter TTVs). Analysis of the TTV of a planet around a single star can provide us additional information about the multi-planetary systems. For example, TTVs for planet pairs near the first order mean motion resonances are well modeled by \citet{lithwick12}, which provided a theoretical method to estimate the masses and eccentricities of planet pairs near the first order mean motion resonances.  A series of   papers \citep{ttvs3, ttvs6, ttvs7, xie12, Yang13} have confirmed more than 60 planets and constrain their masses via TTVs from $Kepler$ data. TTVs of planet pairs in mean motion resonance are also studied by \citet{agol05} and \citet{Boue12}. 

However, TTVs of planets around binary stars are very different from those around single stars. The barycenter of a binary holds still, but the binary stars orbit around each other, and the existence of binary companions will perturb the orbits of planets. The planets around single stars can only be perturbed by additional planets slightly and the transit center times keep linear approximately. However, the TTVs of planets in circumbinary systems are caused by both the gravitational perturbations and the motions of binary stars. Hitherto, seven circumbinary systems have been found by the $Kepler$ space telescope,  i.e., Kepler-16 \citep{Doyle11}, -34, -35 \citep{Welsh12}, -38 \citep{Orosz12b}, -47 \citep{Orosz12a}, -64 \citep{Schwamb13}, and -413 \citep{Kostov14}. Planet transits in these systems show much more complex characteristics in observations (i.e., the duration, depth, and interval of transits) than those around single stars.

In this paper, we focus on planet transits around circumbinary stars. If two or three nearby transits have intervals less than one binary period, we call these nearby transits "tight transits", which occur only in circumbinary systems. We interpret the general sketch of tight transits in Section 2. The characteristics of planetary transit center times, including the intervals and durations of transits as well as the profiles of the light curves are investigated via full three-body integration in Section 3. In Section 4 we interpret how to predict the transits in the next planet period and estimate the occurrence of tight transit. We also study the characteristics of tight transits in the current seven circumbinary systems in Section 5.  Finally, conclusions and some scientific benefits of tight transits are discussed in Section 6. 

\section{Tight Transits in Circumbinary Systems}
Considering that the orbits of the binary and planet are in the same plane, hereafter we denote $a,e,\omega,f,n$ as the semi-major axis, eccentricity, argument of pericenter, anomaly, and angular velocity of a Keplerian orbit, with subscript A,B,P for star A, star B and the planet, respectively. Note that the orbital elements of the planet are relative to the barycenter of the binary, while elements of the binary are relative to the massive star A. The masses of the binary stars are denoted as $m_{\rm A}$ and $m_{\rm B}$ ($m_{\rm A}>m_{\rm B}$). If there is only one transit for each star during one period of the binary orbit, we call it a normal transit similar to transits in single star systems. 

Assuming the stars and a planet are moving in the $X-Z$ plane, and the line of sight is in the direction of $Z$, the transiting planet and the occulted star must have the same projected position $X_{\rm transit}$ in the $X$-axis at the transit center times. The TTVs in circumbinary systems are caused by the variations of $X_{\rm transit}$, which must be less than $2a_{\rm B}$ for star B. Consequently, the TTV for star B must be less than $\frac{2a_{\rm B}}{a_{\rm P}n_{\rm P}}$ if the planet is moving in a circular orbit with the Keplerian angular velocity $n_{\rm P}= 2\pi/ {/rm Peri}_{\rm P}$, where ${\rm Peri}_{\rm P}$ is the period of the planet. Thus, we obtain the TTV of a planet across star B less than $\frac{a_{\rm B}}{\pi a_{\rm P}}{\rm Peri}_{\rm P}$. In the case of Kepler-16, the TTV is less than 22 days, which is much larger than a normal TTV of a planet around single stars. Similiarly, we can also obtain the upper limits of the TTV of a planet across star A. The TTVs in circumbinary systems depend on the semi-major axis ratio of the binary to the planet. Because the more massive binary star A has a smaller semi-major axis, the TTV of a planet across star A is smaller than the less massive binary companion. 

\begin{figure}
\vspace{0cm}\hspace{0cm} \epsscale{1.0} \plotone{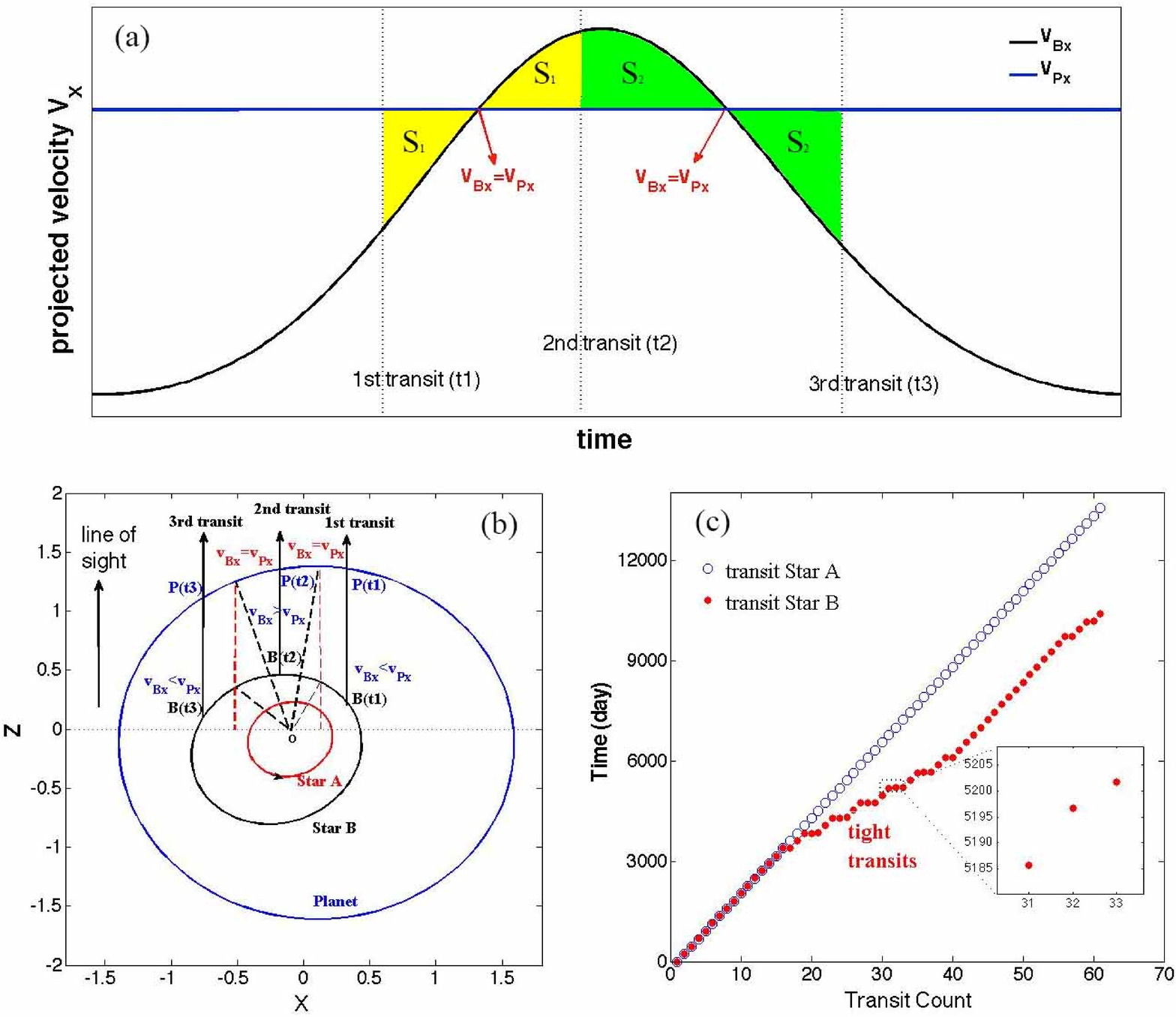} \vspace{0cm} \caption{ (a) Projected velocities of the planet (blue) and star B (black) vary with time in the Kepler-16 system. The vertical dotted lines represent three transits in one binary period. The area of yellow and green regions are $S_1$ and $S_2$, respectively. (b) An intuitive view of tight transits. The three black lines with arrows represent the positions of the transiting planet and occulted star B in three different transits, while the two red dashed lines represent the place where the projected velocity of planet $V_{\rm Px}$ and star B $V_{\rm Bx}$ in the $X$-axis are equal. (c) The transit center time series when the planet transits star A (blue square) and B (red circle) in coplanar Kepler-16 architectures. The  "plateaus" represent tight transits of the planet across star B. \label{fig1}}
\end{figure}

However, a more essential difference is, the planet around circumbinary systems might transit each star more than once. Taking a coplanar Kepler-16 system as an example, Figure \ref{fig1}(a) shows that the projected velocities of the planet and star B in the $X$-axis vary with time, and there can be three transits in one binary period. Obviously, the second transit occurs when the two yellow regions have the same area $S_1$, and the third transit occurs when the two green regions have the same area $S_2$. In the critical case of $S_1=0$ or $S_2=0$, the first or last two transits are combined as one transit, and only two transits occur in one binary period. As shown more intuitively in Figure \ref{fig1}(b), the binary stars A and B are moving in the red and black lines respectively. The velocities of the planet and stars projected to the sky  ($X$-axis) are not the same before the first transit. After the first transit, if star B has a larger projected velocity in the $X$-axis ($V_{\rm Bx}$) than that of the planet ($V_{\rm Px}$), the second transit may occur while star B catches up with the planet. As star B keeps moving, $V_{\rm Bx}$ decreases after the second transit and will change the direction finally. Consequently, the planet must transit star B for the third time. There could be three transits during one binary orbit and we call these transits as "tight transits". This is similar to star A occulted by the planet. 

Adopting a random mean anomaly of planet, we also simulated the transit center times for Kepler-16A and -16B via coplanar three-body integration, as shown in Figure \ref{fig1}(c) . Obviously, tight transits can be found as "plateaus" when star B is occulted in the transit center time series. Different from star B, there are no tight transits while star A is occulted, and we will discuss in detail the necessary condition of the occurrence of tight transits.


The occurrence of tight transits depends on the architecture of the binary and planet. As shown in Figure \ref{fig1}(a), a necessary condition is that the maximum projected velocity of one star is larger than that of the planet, otherwise the star can not catch up with the planet and there is only one transit for each star. Using $\theta_{\rm P}$ (and $\theta_{\rm B}$) to represent the angle between planet (and star B) and the $X$-axis at the transit center time, we have the transit angle of planet $\theta_{\rm P}=\omega_{\rm P}+f_{\rm P}$ and the transit angle of star B $\theta_{\rm B}=\omega_{\rm B}+f_{\rm B}$ . 
The projected velocity in the $X$-axis of star B and the planet are
\beq
 V_{\rm Bx}=\sqrt{\frac{G(m_{\rm A}+m_{\rm B})}{a_{\rm AB}(1-e^2_{\rm AB})}}\mu_{\rm A}(\sin{\theta_{\rm B}}+e_{\rm AB}\sin{\omega_{\rm AB}}),  
\label{VBx}
\eeq
\beq
 V_{\rm Px}=\sqrt{\frac{G(m_{\rm A}+m_{\rm B})}{a_{\rm P}(1-e^2_{\rm P})}}(\sin{\theta_{\rm P}}+e_{\rm P}\sin{\omega_{\rm P}}).
\label{VPx}
\eeq
where $G$ is the gravitational constant and $\mu_{\rm A}=m_{\rm A}/(m_{\rm A}+m_{\rm B})$. $a_{\rm AB}, e_{\rm AB}$, and $\omega_{\rm AB}$ represent the semi-major axis, eccentricity, and argument of pericenter of star B relative to star A. The perturbations on the binary orbit from the planet are very small,  thus we can assume that $a_{\rm AB}, e_{\rm AB}$ and $\omega_{\rm AB}$ are all constants. However, the perturbations from the binary result in variations of planetary orbit with time, as pointed by \citet{LL13}, especially $e_{\rm P}$ and $\omega_{\rm P}$, which obviously depend on time.  It is easily  known that both $V_{\rm Px}$ and $V_{\rm Bx}$ reach the maximum values at $\theta_{\rm P}=\theta_{\rm B}=\pi/2$,  i.e., $X_{\rm P}=X_{\rm B}=0$. So we can obtain a criterion parameter $Q_{\rm ttB}=V_{\rm Bx}/V_{\rm Px}$ for the occurrence of tight transits. In any circumbinary system with the parameter $Q_{\rm ttB}>1$, i.e., if

\beq
Q_{\rm ttB}=\mu_{\rm A}\sqrt{\frac{a_{\rm P}(1-e^2_{\rm P}(t))}{a_{\rm AB}(1-e^2_{\rm AB})}}\frac{1+e_{\rm AB}\sin{\omega_{\rm AB}}}{1+e_{\rm P}(t)\sin{\omega_{\rm P}(t)}}>1,
\label{criterionB1}
\eeq
tight transits can occur. 
We set a time-dependent parameter $z(t)=e_{\rm P}(t)\sin{\omega_{\rm P}}(t)$, and because the eccentricity of planet is always small, we expand $Q_{\rm ttB}$ to the first order of $e_{\rm P}$ and Equation (\ref{criterionB1}) becomes $Q_{\rm ttB}\approx Q_{\rm B0}(1-z(t))>1$. For a given binary system, the tight transit can occur if
\beq
z(t)<1-1/Q_{\rm B0}.
\label{criterionB2}
\eeq
The occurrence of tight transits depends on the secular evolution of the planetary orbit. Using Equation (43) in \citet{LL13}, we can analytically estimate the evolution of $z(t)$ for a given planet around a circumbinary.
%
Equations (\ref{VBx}) and (\ref{criterionB2}) are the same for star A, after replacing the subscripts A and B with each other. Note that $\omega_{\rm BA}=\omega_{\rm AB}+\pi$ and $\mu_{\rm B}=m_{\rm B}/(m_{\rm A}+m_{\rm B})$. Because star A is more massive than star B, i.e., $\mu_{\rm A}>\mu_{\rm B}$, tight transits are more likely to occur for star B with a smaller mass according to Equation (\ref{criterionB1}). Because the secondary star B usual has a low luminosity, transits around star B are not easily detected. However, $Q_{\rm ttB}$ does not depend only on $\mu_{\rm A}$. For some eccentric binary with similar masses, tight transits can also occur and be detected easily.

In systems where tight transits can occur, taking star B as an example, we can calculate the critical angle $\theta_{\rm P}$ and $\theta_{\rm B}$, while $V_{\rm Px}=V_{\rm Bx}$ and $X_{\rm P}=X_{\rm B}$. There must be two different solutions of $\theta_{\rm P}$ and $\theta_{\rm B}$ between [0,$\pi$] by solving equations $X_{\rm P}=X_{\rm B}$ and $V_{\rm Px}=V_{\rm Bx}$, i.e.,
\beq
 \cos{\theta_{\rm P}}=\frac{a_{\rm AB}}{a_{\rm P}}\mu_{\rm A}\frac{(1-e^2_{\rm AB})(1+e_{\rm P}\cos{(\theta_{\rm P}-\omega_{\rm P})})}{(1-e^2_{\rm P})(1+e_{\rm AB}\cos{(\theta_{\rm B}-\omega_{\rm AB})})}\cos{\theta_{\rm B}} ,
\label{XPB}
\eeq
\beq
\sin{\theta_{\rm P}}=\sqrt{\frac{a_{\rm P}(1-e^2_{\rm P})}{a_{\rm AB}(1-e^2_{\rm AB})}}\mu_{\rm A}(\sin{\theta_{\rm B}}+e_{\rm AB}\sin{\omega_{\rm AB}})-e_{\rm P}\sin{\omega_{\rm P}},
\label{Vx}
\eeq
We can easily solve Equations (\ref{XPB}) and (\ref{Vx}) numerically for a given circumbinary system with $e_{\rm P}(t)$ and $\omega_{\rm P}(t)$. The solutions will be time dependent. Setting the two different solutions as $\theta_{\rm P1}$, $\theta_{\rm B1}$ and $\theta_{\rm P2}$, $\theta_{\rm B2}$ (assuming $\theta_{\rm P1}<\theta_{\rm P2}$, $\theta_{\rm B1}<\theta_{\rm B2}$ ), we can conclude that the first transit in tight transits must occur before the planet arrives at $\theta_{\rm P1}$. Otherwise, $V_{\rm B}>V_{\rm P}$ when $\theta_{\rm P}\in[\theta_{\rm P1},\theta_{\rm P2}]$, and the planet can only transit star B after $\theta_{\rm P2}$. Obviously, the second transit can only occur between $\theta_{\rm P1}$ and $\theta_{\rm P2}$, and finally the third transit will occur after $\theta_{\rm P2}$. This characteristics is very important to estimate the interval of tight transits in the next section.

\section{transit counts, durations, intervals and light curves of tight transits}

If a planet is transiting a circumbinary only once during one binary period, it is hard to decide $e_{\rm P}$, $\omega_{\rm P}$, and $\theta_{\rm P}$ at the transit center time accurately. However, if there are three tight transits, we have three durations and two intervals between transits, and there will be much more accurate constraints of the planet orbit. The motivation of this section is to show the influences from $e_{\rm P}$, $\omega_{\rm P}$ and $f_{\rm P}$ on the transit counts, intervals, and durations.

First, we focus on $\theta_{\rm P}$ at the first transit center time. According to Equation (\ref{XPB}), $\theta_{\rm P}$ can be converted to $\theta_{\rm B}$, but instead we choose $\theta_{\rm B}(t1)$ as the variable, which is the angle of $\theta_{\rm B}$ when the first transit occurs. The time dependency of $\theta_{\rm B}(t1)$ will be discussed in Section 4. In this section, we will interpret how the transits go from being normal to being tight transits by scanning $\theta_{\rm B}(t1)$, while the initial value of other five orbital elements of planet are set as observational values. Our coplanar three-body simulation results are shown in Figure \ref{fig2}(a). The four black dashed lines denote four characteristic angles $\theta'_{\rm B3}$, $\theta''_{\rm B3}$, $\theta'_{\rm B1}$, and $\theta''_{\rm B1}$, which represent the boundaries of $\theta_{\rm B}(t1)$ when $N_{\rm transit}$ begins to change. The two red dashed lines represent the two critical cases according to the two solutions of Equations (\ref{XPB}) and (\ref{Vx}) with the assumption that the orbital elements of the planet are equal to their initial values. 

In Figure \ref{fig2}(a), when $\theta_{\rm B}(t1)$ is small, star B can not catch up with the planet in $X-$direction; therefore, only one transit occurs, i.e., a normal transit. If $\theta_{\rm B}(t1)=\theta'_{\rm B3}$, the second transit will occur around $\theta_{\rm B}\approx \theta_{\rm B2}$, where $V_{\rm Px} \approx V_{\rm Bx}$. However,  in this case, the planet can only overlap with the edge of star B; therefore, the duration is very short. When $\theta_{\rm B}(t1)$ increases, the planet can transit the center of the star and even go across the stellar surface. Therefore, the distance traveled by the planet during this transit increases and thus the duration becomes longer and longer, as shown in Figure \ref{fig2}(b). After the planet goes across the stellar surface at the second time, the projected velocity of star B in the $X$-axis will move in the opposite direction, and consequently a third transit occurs when the planet catches up with star B. 

The region between $\theta'_{\rm B3}$ and $\theta''_{\rm B3}$ is chaotic because of the perturbation of binary gravities. The velocity of the planet is perturbed by binary stars, and a tiny variation of $V_{\rm Px}$ can lead to different $N_{\rm transit}$ in this critical region, where $V_{\rm Px}\approx V_{\rm Bx}$ during the second transit. Another region between $\theta'_{\rm B1}$ and $\theta''_{\rm B1}$ is similar. Therefore, it is hard to decide $N_{\rm transit}$ (2 or 3) in these two chaotic regions. As shown in Figures \ref{fig2}(b) and (c), the duration of the second transit when $N_{\rm transit}$=2 is twice as long as the duration of the second transit when $N_{\rm transit}$=3. In Figure \ref{fig2}(c), we consider the second transit when $N_{\rm transit}$=2 as a combination of two bifurcated transits. The duration has a negative correlation with the size of the overlapped region for the two combined transits. The two transits are separated when $\theta_{\rm B}(t1)=\theta'_{\rm B1}$ or $\theta''_{\rm B1}$ and are overlapped totally when $\theta_{\rm B}(t1)=\theta_{\rm B1}$. So the duration has a minimum value when $\theta_{\rm B}(t1)=60^{\rm o}$.29 as shown in Figure \ref{fig2}(c). When tight transits occur in a circumbinary system, the smallest transit angle $\theta'_{\rm B3}$ corresponds to $S_2=0$ in Figure \ref{fig1}(a), i.e., the first transit occurs at the earliest time. Similarly, if $S_1=0$, we can also obtain the largest critical angle $\theta''_{\rm B4}$, which corresponds the last transits that occur at the latest time.  The relative phases of the planet and the star have to line up just right with $\theta_{\rm B}\in [\theta'_{\rm B3},\theta''_{\rm B4}]$ at the transit center time, so that tight transits can occur in the current planetary period.

We do not discuss the cases when $\theta_{\rm B}(t1)>\theta''_{\rm B1}$, because the first transit must occur while $\theta_{\rm B}(t1)\le \theta''_{\rm B1}$. In the following, we set the first, second, and third transit center times as $t1$, $t2$, and $t3$ in tight transits, and the corresponding angles are $\theta_{\rm P}(t1),\theta_{\rm P}(t2)$, and $\theta_{\rm P}(t3)$ for planets and $\theta_{\rm B}(t1),\theta_{\rm B}(t2), and \theta_{\rm B}(t3)$ for star B.

The duration of a transit $\approx \frac{1}{|V_{\rm Px}-V_{\rm Bx}|}$.
At the beginning when $N_{\rm transit}$=1, ${|V_{\rm Px}-V_{\rm Bx}|}$ decreases with $\theta_{\rm B}(t1)$ before $\theta'_{\rm B1}$, hence the duration of the transit increases. While $N_{\rm transit}$=2, the second transit ($\theta_{\rm B}(t2)\in [\theta'_{\rm B3},\theta''_{\rm B3}]$) or the first transit ($\theta_{\rm B}(t1)\in [\theta'_{\rm B1},\theta''_{\rm B1}]$) occurs where $V_{\rm Px}\approx V_{\rm Bx}$, and the duration can become as long as $\sim$4 days. When $N_{\rm transit}$=3, the transit with the longest duration is bifurcate as two separated transits with a shorter duration than $N_{\rm transit}$=2. In the case of Kepler-16, the duration of the second transit is longer than the other two when $N_{\rm transit}$=3. However, the durations of tight transits depend on the architectures of binary and planetary orbits in general cases. 


As the definition of tight transits, all transits must occur in one binary period, ${\rm Peri}_{\rm AB}$. Meanwhile, as mentioned in Section 2, the first transit occurs before $\theta_{P1}$, and the last one occurs after $\theta_{\rm P2}$.  Therefore, the interval of the tight transits $(t3-t1)$ satisfies
\beq
{\rm Peri}_{\rm AB}>t3-t1>\frac{\theta_{\rm P2} - \theta_{\rm P1}}{2\pi}{\rm Peri}_{\rm P}.
\label{interval}
\eeq
The lower limit of ($t3-t1$) depends on the solution of Equations (\ref{XPB}) and (\ref{Vx}). Note that in Equation (\ref{interval}), we assume the planet is in a circular orbit. The blue dotted line in Figure \ref{fig2}(a) shows the lower limit during tight transits in current Kepler-16 architectures.

To study the profiles of light curves in tight transits, we choose eight different values of  $\theta_{\rm P}(t1)$ as noted in Figure \ref{fig3}, covering different cases with $N_{\rm transit}$=1,2, and 3. The limb darkening effect of Kepler-16B is considered to be the same as \citet{Mandel02}, adopting the coefficients $c_2=0.6$ and $c_1=c_3=c_4=0$. The light curves in these cases and the intervals between them are shown in Figure \ref{fig3}. The intervals are calculated as $T_{\rm in}-T_{\rm out}$, where $T_{\rm out}$ is the time when the former transit ends and $T_{\rm in}$ is the time when the later transit starts. In the two narrow middle panels of Figure \ref{fig3}, intervals of eight cases are plotted as circles from top to bottom.

In the case of  $\theta_{\rm B}(t1)=27^{\rm o}.4$, the light curve of the second transit has a smaller depth than others, because the impact parameter $b$ changes from 1 to 0.61 and then increases to 1, i.e., the planet does not go across the center of star B. When $\theta_{\rm B}(t1)=28^{\rm o}.02$, the duration of the second transit is $\sim$3 days, much larger than the duration of the first normal transit of $\sim0.4$ days. The durations of "ingress" and "egress" are $\sim$0.5 day, much larger than normal values. There is a bifurcation at $\theta_{\rm B}(t1)=29^{\rm o}.10$, and the interval between the second and the third transit is very small ($\sim0.3314$ day). In this critical case, the asymmetries of the light curves of the last two transits are obvious.  The "egress" of the second transit and "ingress" of the third transit are elongated. As we increase $\theta_{\rm B}(t1)$, the interval between the last two transits increases while the interval between the first two transits changes contrarily. Another bifurcation comes out at $\theta_{\rm B}(t1)=51^{\rm o}.38$.

Considering the variations of $e_{\rm P}$ and $\omega_{\rm P}$ in a long timescale, the solutions of Equations (\ref{XPB}) and (\ref{Vx}) change, i.e., the same $\theta_{\rm B}(t1)$ will lead to different characteristics of transits due to the different characteristic angles $\theta'_{\rm B3}$, $\theta''_{\rm B3}$, $\theta'_{\rm B1}$, and $\theta''_{\rm B1}$. However, during one observation of tight transits shorter than ${\rm Peri}_{\rm AB}$, we can also assume that $e_{\rm P}$ and $\omega_{\rm P}$ are independent of time, i.e., the four characteristic angles remain constant. The analysis above is still applicable by changing the values of four characteristic angles. 

\section{time dependency and occurrence of tight transit}
We have investigated some critical characteristics of tight transits. Another interesting question is how can we predict tight transits based on an observed transit? In this section, we will discuss the criterion of a tight transit in the next planet period based on the previous $\theta_{\rm B}(t1)$. 

Assuming a transit is observed at the current epoch, we can obtain the transit angle $\theta_{\rm B0}$ at the observed transit center time. As pointed in Section 3, if $\theta_{\rm B0}\in [\theta'_{\rm B3},\theta''_{\rm B4}]$, tight transits will occur in the current binary period. We set $\theta_{\rm P0}$ as the transit angles of the planet while the first transit occurs. $T_{1}$ represents the remainder of ${\rm Peri}_{\rm P}$ divided by ${\rm Peri}_{\rm AB}$, where $T_{1}\in (-{\rm Peri}_{\rm AB}/2,{\rm Peri}_{\rm AB}/2]$, and ${\rm Peri}_{\rm P}$ is the mean period of the next Keplerian orbit, i.e., after ${\rm Peri}_{\rm P}$, the planet still has the same position with the angle of $\theta_{\rm P0}$. Similarly, star B has the same position with the angle of $\theta_{\rm B0}$ after several ${\rm Peri}_{\rm AB}$. In the next period of the planet, when transit occurs the transit angles of the planet $\theta_{\rm P1}$ and star B $\theta_{\rm B1}$ must satisfy
\beq
\int^{\theta_{\rm P1}}_{\theta_{\rm P0}} \frac{d\theta_{\rm P}}{\dot{\theta}_{\rm P}}+T_{1}=\int^{\theta_{\rm B1}}_{\theta_{\rm B0}} \frac{d\theta_{\rm B}}{\dot{\theta}_{\rm B}}.
\label{BP1}
\eeq

The perturbation from the planet on a binary orbit can be ignored; therefore, $\dot{\theta}_{\rm B}=n_{\rm AB}(1+e_{\rm AB}\cos(\theta_{\rm B}-\omega_{\rm AB}))^2(1-e^2_{\rm AB})^{1/2}$, where $n_{\rm AB}$ is the Keplerian angular velocity of the binary. $\dot{\theta}_{\rm P}=\dot{f}_{\rm P}+\dot{\omega}_{\rm P}$, the perturbation of the binary, makes $\dot{\theta}_{\rm P}$ different from Keplerian motion. One of the most complex challenges is to derive an analytical expression of $\dot{\theta}_{\rm P}$, which depends on the initial binary orbital phase $\theta_{\rm B0}$. A much easier way to obtain accurate $\dot{\theta}_{\rm P}$ is by performing three-body simulations. Combining Equations (\ref{XPB}) and (\ref{BP1}), we can solve $\theta_{\rm B1}$ and $\theta_{\rm P1}$. If $\theta_{\rm B1}\in [\theta'_{\rm B3},\theta''_{\rm B4}]$, the tight transits will occur in the next period of the planet.

If $T_1=0$, we can easily obtain $\theta_{\rm B1}=\theta_{\rm B0}$ according to Equations (\ref{XPB}) and (\ref{BP1}). Generally, we can always obtain an integer $j$, which makes $T_j=j*{\rm Peri}_{\rm P}$ mod ${\rm Peri}_{\rm AB}$ small, i.e., $T_j/{\rm Peri}_{\rm AB}\sim 0$. In the case of $T_j/{\rm Peri}_{\rm AB}\sim0$. We set a small value of $d\theta_{\rm B}=\theta_{\rm B1}-\theta_{\rm B0}$, and Equation (\ref{BP1}) becomes
\beq
\frac{d\theta_{\rm P}}{\dot{\theta}_{\rm P}}+T_j=\frac{d\theta_{\rm B}}{\dot{\theta}_{\rm B}}.
\label{BP2}
\eeq
Noting that $d\theta_{\rm B}$ and $d\theta_{\rm P}$ are therefore small in such a short timescale, we can adopt
\beq
\dot{\theta}_{\rm B}\approx n_{\rm AB} (1+2e_{\rm AB}\cos(\theta_{\rm B}-\omega_{\rm AB}))+O(e^2_{\rm AB}),
\eeq 
\beq
\dot{\theta}_{\rm P}\approx n_{\rm P} (1+2e_{\rm P}\cos(\theta_{\rm P}-\omega_{\rm P}))+O(e^2_{\rm P}).
\eeq
The differentiation of Equation (\ref{XPB}) is
\beq
\begin{split}
&-\mu_{\rm A}a_{\rm AB}(1-e^2_{\rm AB})\frac{\sin\theta_{\rm B}+e_{\rm AB}\sin\omega_{\rm AB}}{(1+e_{\rm AB}\cos(\theta_{\rm B}-\omega_{\rm AB}))^2}d\theta_{\rm B}\\
&=-a_{\rm P}(1-e^2_{\rm P})\frac{\sin\theta_{\rm P}+e_{\rm P}\sin\omega_{\rm P}}{(1+e_{\rm P}\cos(\theta_{\rm P}-\omega_{\rm P}))^2}d\theta_{\rm P}
\end{split}
\label{dXPB}
\eeq

Note that Equations (\ref{BP2})-(\ref{dXPB}) are available only if $T_j/{\rm Peri}_{\rm AB}$ is small. Assuming the planets in circumbinary systems have small eccentricities, we ignore the terms containing $e_{\rm P}$ and only reserve the first order terms of $e_{\rm AB}=0$. Meanwhile, $\theta_{\rm P}\sim90^{\rm o}$ when the planet transits each binary star. Approximately, we can obtain
\beq
d\theta_{\rm B}=\frac{2\pi\cdot(T_j/{\rm Peri}_{\rm AB})(1+2e_{\rm AB}\cos(\theta_{\rm B}-\omega_{\rm AB}))}{1-\mu_{\rm A}(\frac{{\rm Peri}_{\rm P}}{{\rm Peri}_{\rm AB}})^{1/3}(\sin\theta_{\rm B}+e_{\rm AB}\sin\omega_{\rm AB})} 
\label{dtheta}
\eeq
Equation (\ref{dtheta}) is invalid when the denominator is small, because $d\theta_{\rm B}$ becomes large in such a condition. In Kepler-16 systems, taking $j=2$, we have $T_2/{\rm Peri}_{\rm AB}=2{\rm Peri}_{\rm P}/{\rm Peri}_{\rm AB}=0.137$, which means $\theta_{\rm B}$ changes to $\theta_{\rm B}+d\theta_{\rm B}$ every two planetary periods. This is why there are several tight transits that occur every two ${\rm Peri}_{\rm P}$ in Figure \ref{fig1}(c).

The possibility of a transit with transit angle $\theta_{\rm B}$ is $p(\theta_{\rm B})\propto1/|d\theta_{\rm B}|$, i.e., 
\beq
p(\theta_{\rm B})\propto |\frac{1-\mu_{\rm A}(\frac{{\rm Peri}_{\rm P}}{{\rm Peri}_{\rm AB}})^{1/3}(\sin\theta_{\rm B}+e_{\rm AB}\sin\omega_{\rm AB})}{1+2e_{\rm AB}\cos(\theta_{\rm B}-\omega_{\rm AB})}|
\label{ptransit}
\eeq
Neglecting the terms of $e_{\rm AB}$, Equation (\ref{ptransit}) is simplified as
\beq 
p(\theta_{\rm B})\propto |1-\mu_{\rm A}(\frac{{\rm Peri}_{\rm P}}{{\rm Peri}_{\rm AB}})^{1/3}\sin\theta_{\rm B}|.
\label{ptransit1}
\eeq
Here we must emphasize that Equation (\ref{ptransit1}) is only available when both $e_{\rm AB}$ and $e_{\rm P}$ are small.

According to Equation (\ref{ptransit1}), the possibility becomes minimum or maximum when $\theta_{\rm B}=90^{\rm o}$ or $270^{\rm o}$, respectively. Since tight transits will occur when $\theta_{\rm B}\in [\theta'_{\rm B3},\theta''_{\rm B4}]$, the occurrence rate of tight transits must be smaller than $\frac{\theta''_{\rm B4}-\theta'_{\rm B3}}{2\pi}$ because the possibility of $\theta_{\rm B}$ is sinusoidal. Using a coplanar three-body model, we simulate the transits in Kepler-16, -34, and 35 systems in $10^4$ yr (Kepler-38, -47, and -64 systems will be discussed in Section 5). The fraction of tight transits can be obtained, and we can also count the possibility of different transit angles $\theta_{\rm B}$. Figure \ref{fig4} shows the possibilities of different $\theta_{\rm B}$ in Kepler-16, -34, and -35 systems. Obviously, the distributions in Kepler-16 and -35 systems correspond with our estimated probability in Equation (\ref{ptransit1}). The occurrence rates of tight transits $f_{\rm tt}$ in Kepler-16 and -35 systems are $7.1\%$ and $2.9\%$, respectively. Because of $e_{\rm AB}=0.52$ in the Kepler-34 system, it is not suitable to estimate $p(\theta_{\rm B})$ via Equation (\ref{ptransit1}), especially when $p(\theta_{\rm B})\sim0$. Based on our simulation, $f_{\rm tt}=11.2\%$, i.e., $11.6\%$ transits events in Kepler-34 are tight transits. It is easier to detect the tight transits in Kepler-16 and -34 systems rather than in Kepler-35 systems.   

\section{Current seven Circumbinary systems}
Currently, seven circumbinary systems have been detected by the $Kepler$ space telescope, and only a few transits are observed for each planet in these systems. The former studies are all based on a coplanar model. In this section, we consider the real architecture of each system and check if the tight transits for each star can occur in these systems by three-body integrations. The mutual inclination between the binary and planet is $\Delta i \sim 0.01$. The influences on the projected velocity due to the mutual inclination is as small as $\cos(\pi/2)-\cos(\pi/2-\Delta i) \sim \Delta i^2$. However, mutual inclination will decide whether the planet can transit the binary stars or not.

The criterion parameters $Q_{\rm ttA}$ and $Q_{\rm ttB}$ are listed in Table \ref{tbl-1}. We also simulate the maximum $|z|$ in 100 yr due to perturbations of the binary. Tight transits for the massive star A in all seven systems are forbidden because $Q_{\rm ttA}<1$ even when $z=-|z|_{\rm max}$. Setting $z=|z|_{\rm max}$, all the other six systems with $Q_{\rm ttB}>1$ allow the occurrence of tight transits for the less massive star B, except Kepler-413. However, considering the mutual inclination between the planet and star B, planets in Kepler-38 and -64 can not transit star B because the projected orbits of the planet and star B do not overlap. Therefore, tight transits can occur in the real Kepler-16, -34, -35, and -47 systems. The corresponding solutions of Equations (\ref{XPB}) and (\ref{Vx}) for each system  with $e_{\rm P0}$ and $\omega_{\rm P0}$ are also listed in Table \ref{tbl-1} as a reference of characteristic angles. When the transit angle $\theta_{\rm B}(t1)$ is in the range of [$\theta'_{\rm B3}, \theta''_{\rm B1}$], the tight transits must occur.  It is easy to understand that the larger $Q_{\rm ttB}$ is, the larger the range of $[\theta_{\rm B1},\theta_{\rm B1}]$ is, because the $Q_{\rm ttB}$ represents the degree of difficulty of tight transits. Meanwhile, Kepler-47c has the largest region from $-56^{\rm o}.38$ to $38^{\rm o}.32$ for tight transit due to the largest $Q_{\rm ttB}=2.5221$ with the current architecture. We also predict when the most recent tight transits begin in these systems in the last row in Table \ref{tbl-1}. Assuming the eccentricities of two planets are 0, Kepler-47 will have an earliest tight transits while Kepler-16, -34, and -35 need to wait quite a long time for the first tight transits.

In the real Kepler-47 system, $e_{\rm P}$ and $\omega_{\rm P}$ of the two planets are not determined, and we only have $e_{\rm P}<0.035$ for Kepler-47b and $e_{\rm P}<0.411$ for Kepler-47c. Setting $e_{\rm P}=0$ and $\omega_{\rm P}=0$ for Kepler-47b, we scanned $e_{\rm P}\sin(\omega_{\rm P})$ and $e_{\rm P}\cos(\omega_{\rm P})$ of Kepler-47c and estimate the most recent date that tight transits occur via full four-body integrations. The most recent date for Kepler-47b and -47c are shown in Figure \ref{fig5}, although the transits around the faint star Kepler-47B are not easily detected. In most cases, tight transits will occur before 2024 for both Kepler-47b and -47c. In particular, Kepler-47c can experience tight transits before 2016 in some cases. Only if the eccentricity of Kepler-47c is larger than 0.25, the tight transits likely occur after 2024. Note that the large purple region in the left panel of Figure \ref{fig5} does not mean there are no differences. In most cases of the purple region, the most tight transits begin on 2020 September 15, with a small difference less than 9 hr. It indicates that the small eccentricity of Kepler-47c can only influence the tight transits of Kepler-47b to a limited degree, which is still detectable. It is good for us to focus on Kepler-47 systems over the next 2 yr, since this will probably provide us more constraints on this system via observed characteristics of tight transits.


\section{Discussions and Conclusions }
\begin{figure}
\vspace{0cm}\hspace{0cm} \epsscale{1.0} \plotone{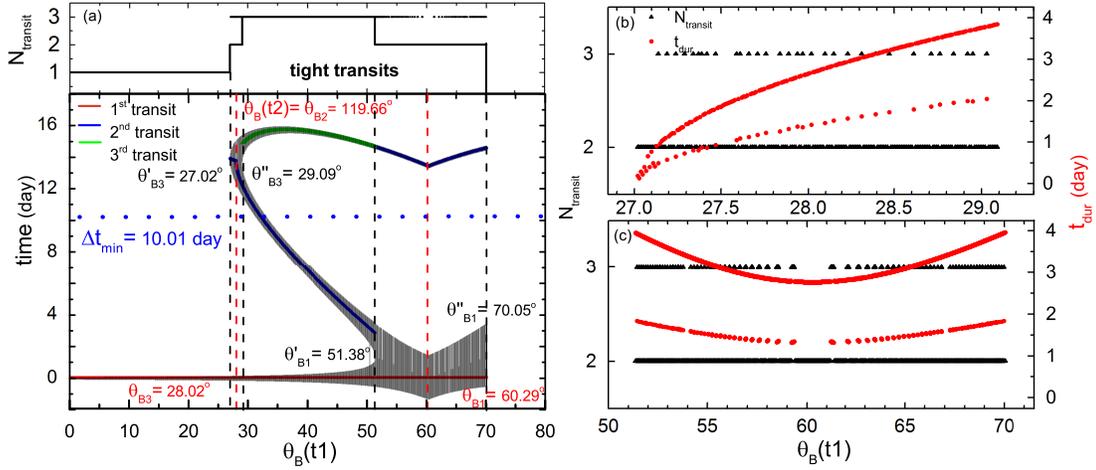} \vspace{0cm} \caption{(a) Transit counts ($N_{\rm transit}$), transit center time (time), and duration ($t_{\rm dur}$) of coplanar Kepler-16 architectures are varying with $\theta_{\rm B}(t1)$. The transit counts are plotted on the top left panel. If $N_{\rm transit}>1$, tight transits occur. The red, black, and blue lines represent the transit center time of the first, second and third transits, respectively. The durations of each transit are plotted as the gray region. The four black dashed lines represent the boundaries of $\theta_{\rm B}(t1)$ where $N_{\rm transit}$ begin to change while the two red dashed lines represent the two critical values $\theta_{\rm B3}$ and $\theta_{\rm B1}$. (b) The enlarged plot between $\theta'_{\rm B3}$ and $\theta''_{\rm B3}$. The duration of the second transit when $N_{\rm transit}=3$ becomes half of that when $N_{\rm transit}=2$. (c) The same with panel (b), but in the region from $\theta'_{\rm B1}$ to $\theta''_{\rm B1}$. \label{fig2}}
\end{figure}

\begin{figure}
\vspace{0cm}\hspace{0cm} \epsscale{1.0} \plotone{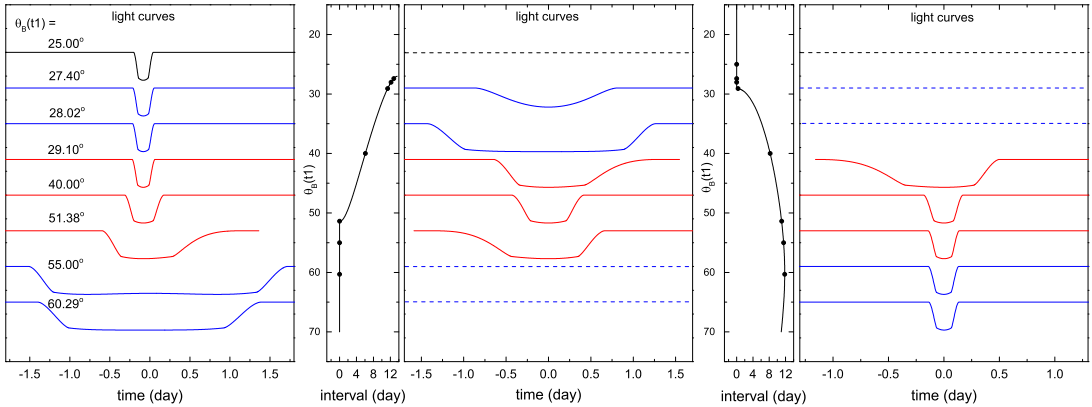} \vspace{0cm} \caption{Light curves of transits of coplanar Kepler-16 architectures when $\theta_{\rm B}(t1)=25^{\rm o}.00$, $27^{\rm o}.40$, $28^{\rm o}.02$, $29^{\rm o}.10$, $40^{\rm o}.00$, $51^{\rm o}.38$, $55^{\rm o}.00$, and $60^{\rm o}.29$ from top to bottom, respectively. Dashed horizontal lines mean there is no transit and the light curves are exactly constant. The cases of $N_{\rm transit}$=1, 2, and 3 are distinguished by black, blue, and red lines. The intervals of nearby transits are also plotted in the two narrow middle panels between light curves, and the eight black dots correspond to the eight cases from top to bottom.\label{fig3}}
\end{figure}

\begin{figure}
\vspace{0cm}\hspace{0cm} \epsscale{1.0} \plotone{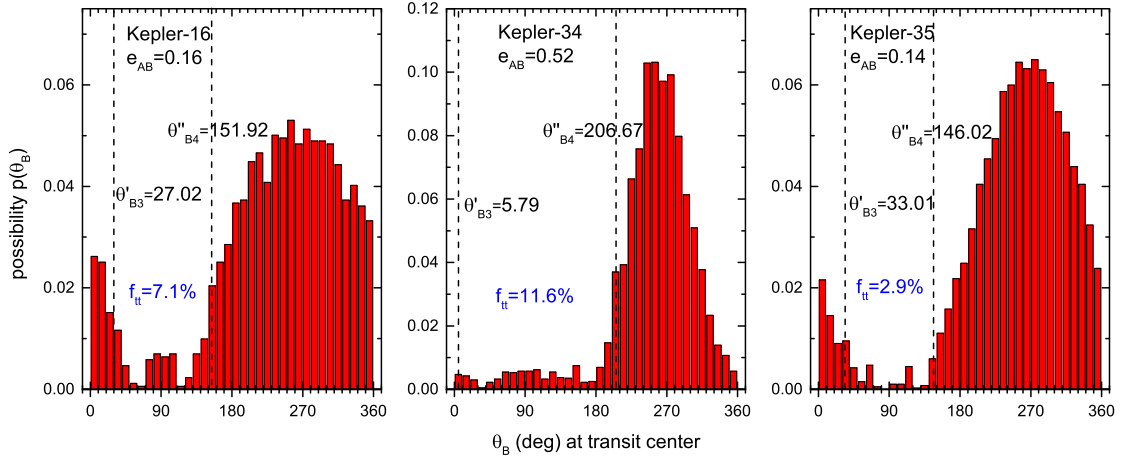} \vspace{0cm} \caption{ Possibilities of different transit angles $\theta_{\rm B}$ in coplanar Kepler-16, -34, and -35 systems. The occurrence fraction of tight transit $f_{\rm tt}$ is $7.1\%$, $11.6\%$, and $2.9\%$ in these three systems, respectively. Note that the possibilities are sinusoidal rather than uniform; therefore, $f_{\rm tt}$ is much smaller than $\frac{\theta''_{\rm B4}-\theta'_{\rm B3}}{2\pi}$. The profiles of $p(\theta_{\rm B})$ correspond with Equation (\ref{ptransit1}), except the Kepler-34 system because of the large $e_{\rm AB}=0.52$. \label{fig4}}
\end{figure}

\begin{figure}
\vspace{0cm}\hspace{0cm} \epsscale{1.0} \plotone{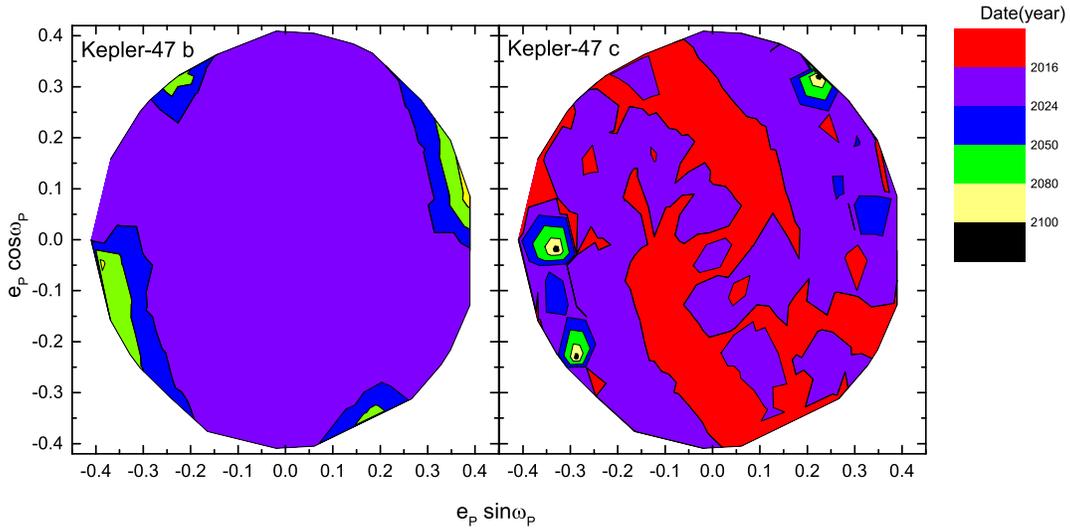} \vspace{0cm} \caption{ Most recent data for when tight transits of Kepler-47b (left) and -47c(right) occur. In our full four-body integrations, the eccentricity and the argument of pericenter of Kepler-47b are fixed as 0 according to the observational limitation of $e_{\rm P}<0.035$. Meanwhile $e_{\rm P}\sin(\omega_{\rm P})$ and $e_{\rm P}\cos(\omega_{\rm P})$ of Kepler-47c are scanned, with the observational limitation of $e_{\rm P}<0.411$. Our results show that Kepler-47c can experience tight transits before 2016; meanwhile, tight transits will occur between 2016 and 2024 for Kepler-47b in most cases. Note that in most cases of the purple region in the left panel, the most recent tight transits occur on 2020 September 15, with a small variation ($<$ 9 hr) due to different eccentricities and $\omega$ of Kepler-47c. These small differences are not represented in this figure but are detectable. \label{fig5}}
\end{figure}

We use a coplanar Kepler-16 circumbinary system to exemplify the occurrence of tight transits in general circumbinary systems in Section 2. We obtain a necessary criterion to judge when tight transits can occur in a circumbinary system, as shown in Equation (\ref{criterionB2}). If tight transits can occur, there are two bifurcations near the regions where the projected velocities of the planet and star B in the $X$-axis are equal. The transit count changes at the two bifurcations. The durations, light curves and interval between tight transits in coplanar Kepler-16 system are shown in Figures \ref{fig2} and \ref{fig3}.  As shown in Equation (\ref{interval}), the interval of the tight transits must be less than the period of the binary and larger than a variable minimum interval. The durations of tight transits can become much longer than a normal transit, thus tight transits can be detected easily. Using the observational data in tight transits will improve the accuracy of the planetary orbit. The tight transits are not likely to occur continuously unless ${\rm Peri}_{\rm P}\sim j\cdot {\rm Peri}_{\rm AB}$. As shown in Equation (\ref{ptransit1}), we can estimate the occurrence of tight transits by integral $\theta_{\rm B}$ from $\theta'_{\rm B3}$ to $\theta''_{\rm B1}$. Figure \ref{fig4} shows the distributions of $\theta_{\rm B}$ at transit center times in $10^4$ yr in Kepler-16, -34, and -35 systems as well as the occurrence rate of tight transits. We have investigated the tight transits in the current seven circumbinary systems. Some orbital parameters of seven detected circumbinary systems are listed in Table \ref{tbl-1}, as well as some critical angles and criterion parameters. Considering the mutual inclinations between the planet and the binary, tight transits can occur when Kepler-16b, -34b, -35b, and -47b transit their own star B. We also predict the occurrence time of the most recent tight transits in these systems in the next 100 yr, which are listed in the last row of Table \ref{tbl-1}. Although the eccentricity of Kepler-47c is uncertain, it is probable that tight transits will occur before 2024 for both Kepler-47b and -47c, according to Figure \ref{fig5}.   

Tight transits provide more constraints to the architectures of binary and planetary orbits. If tight transits are observed, a good way to minimize the error of $e_{\rm P}, \omega_{\rm P}$ and $f_{\rm P}$ is by fitting the transit properties. As the Earth-like planets can be formed in circumbinary systems \citep{Gong13} and habitable Earths are stable in some of these systems \citep{Liu13}, the additional habitable Earth in these systems will perturb the orbits of the current planets. Although the perturbation is small, the differences of  transit time can be enlarged and become observable during tight transits. Similarly, perturbation from an exomoon is also possible to be detected during tight transits. The duration of about 3 days also makes the detection of an exomoon easy based on the simulation of transit light curves by \citet{Kipping09} and \citet{Simon09}. 

As pointed out in Section 3,  the tight transits can have a long duration or "ingress/egress" during one transit. With a fixed observation cadence,  we can obtain more data during a single transit. The long "ingress/egress" during tight transits allow us to detect the oblateness of planets that can be used to constrain the rotation rate of the planet \citep{Seager02,Winn09,Rag09}. Starspots and limb darkening effects are also suitable for investigate during tight transit because the planets go through the star surface much slower than a normal transit and more observation data can be obtained during tight transits. 


We thank Dr. Darin Ragozzine for providing helpful comments. This work is supported by National Basic Research Program of China (2013CB834900), National Natural Science Foundations of China (Nos.10925313 and 11333002), the Strategic Priority Research Program-The Emergence of Cosmological Structures of the Chinese Academy of Sciences (grant No. XDB09000000), the Natural Science Foundation for the Youth of Jiangsu Province (No. BK20130547), the 985 project of Nanjing University, the Superiority Discipline Construction Project of Jiangsu Province, and the National Natural Science Funds for Young Scholar(No. 11003010).



\begin{table}
\begin{center}
\caption{Orbital Parameters and Critical Angles of Tight Transits in the Seven Circumbinary Systems. \label{tbl-1}}.
\begin{tabular}{lcccccccc}
\hline
\\ 
        & Kepler-16b  & Kepler-34b  & Kepler-35b & Kepler-38b & Kepler-47b  & Kepler-47c  & Kepler-64b  & Kepler-413b\footnote{Data of binary stars and planets used in this table are all from \citet{Doyle11}, \citet{Welsh12}, \citet{Orosz12b}, \citet{Orosz12b}, \citet{Orosz12a}, \citet{Schwamb13}, and \citet{Kostov14}, respectively.} \\
\hline
${\rm Peri}_{\rm AB}$(day) &   41.079220    & 27.7958103    & 20.733667       & 18.79537      & 7.44837695   &  7.44837695  &  20.000214  & 10.1161114   \\
${\rm Peri}_{\rm P}$(day)   &   228.776        & 288.822          & 131.455           & 105.595        & 49.514           &303.158           &  138.506     &  66.262  \\
$e_{\rm AB}$		&  0.15944          &   0.52      & 0.1421    &  0.1032   &  0.0234&  0.0234  &  0.2165   &  0.0365 \\
$e_{\rm P0}$			&  0.0069            &   0.182    & 0.042     &  0.016     &  0.0       &  0.0        &  0.0539   &  0.1181 \\
$\omega_{\rm AB}$	&  263.464          &   71.4       &   82.276   &    268.68     &   212.3   &  212.3        &  217.6     &  279.54  \\
$\omega_{\rm P0}$(deg)    &  318         &    7.9       & 63.43     &  200        &   0.0   &  0.0         &  348     &  94.6  \\

\hline
$Q_{\rm ttA}$   \footnote{$Q_{\rm ttA}$ and $Q_{\rm ttB}$ is calculated via Equation (\ref{criterionB1}), $z=e_{\rm P}(t)\sin{\omega_{\rm P}(t)}$.}	   & 0.4722($1-z$)         & 0.6386($1-z$)       &  0.7669($1-z$)       & 0.4099($1-z$)        & 0.4907($1-z$)  &  0.8975($1-z$)  &  0.4822($1-z$)  & 0.7722($1-z$)\\
$Q_{\rm ttB}$         & 1.1681($1-z$)         & 1.9343($1-z$)        &  1.1146($1-z$)       & 1.2699($1-z$)          &  1.3788($1-z$)    &  2.5221($1-z$)  &  1.3253($1-z$)  & 1.0865($1-z$) \\
$z=$(now)            & 0.0046 & 0.025 & 0.037 & -0.0055 & 0.0 & 0.0 & 0.011  &  0.1177 \\
$|z|_{\rm max}$ \footnote{The maximum $|z|$ in 100 yr due to dynamical secular evolution via integration.}   & 0.09 & 0.23 & 0.08 & 0.05 & 0.028 & 0.0024  & 0.142   &   0.15     \\
${\rm flag}_{\rm tB}$ \footnote{${\rm Flag}_{\rm tB}$=1 (or 0) means the projected
orbits of the planet and star B are (or are not) overlapped considering the evolution of mutual inclination in 100 yr, i.e., the planet can (or can not) transit star B in the current architecture in 100 yr.}           &   1    &  1   &  1    &  0     &  1   &  1  &  0  & 1   \\
\hline
$\theta_{\rm B1} $(deg)\footnote{All these angles are calculated using the current $e_{\rm P0}$ and $\omega_{\rm P0}$}    
							&  60.29     & 18.05    &  67.02     &   52.77    &  46.11    &    23.79    &  50.49   & -   \\
$\theta_{\rm B2} $(deg)           & 119.66    & 162.01  &112.98     & 127.23    & 133.85   &  156.21   & 129.17   & -   \\
$\theta_{\rm P1} $(deg)           &   82.03    & 86.55    &  86.93     &   80.65    &   81.44   &    86.63    &  80.48   & -   \\
$\theta_{\rm P2} $(deg)           &   97.78    & 94.44    &  93.10     &   99.37    &   98.32   &    93.25    &  97.29   & -   \\
$\theta'_{\rm B3} $(deg)          &   27.02    &   5.79    &  33.01     &   10.59    &   -2.13   &  -56.38     &  11.39   & -  \\
$\theta''_{\rm B4} $(deg)          & 151.92    &206.67   &146.02     & 167.62    &  181.22  & 237.82      & 181.15  & - \\
\hline
  Date\footnote{The time of the most recent tight transits of planet across star B in 100 yr by three-body integration fully.}       & 2089 Jan 29  & 2048 Apr 12  & $>${\rm 2114 year} & ${\rm Flag}_{\rm tB}=0$ & 2020 Sep 18  & 2021 Jan 19  & $ {\rm Flag}_{\rm tB}=0$ &  Forbidden \\
\hline
\end{tabular}
\end{center}
\end{table}

\begin{thebibliography}{99}
\bibitem[Agol et al.(2005)]{agol05} Agol, E., Steffen, J., Sari, R., \& Clarkson, W.\ 2005, \mnras, 359, 567
\bibitem[Bou{\'e} et al.(2012)]{Boue12} Bou{\'e}, G., Oshagh, M., Montalto, M., \& Santos, N.~C.\ 2012, \mnras, 422, L57
\bibitem[{Doyle et al.,(2011)}]{Doyle11}Doyle, L. R., Carter, J. A., Fabrycky, D. C., et al. \ 2011, Sci, 333, 1602
\bibitem[{Gong et al.,(2013)}]{Gong13}Gong, Y.-X., Zhou, J.-L., \& Xie, J.-W. \ 2013, ApJL, 763, L8
\bibitem[Kipping et al. (2009)] {Kipping09} Kipping, D. M., Fossey, S. J. \& Campanella, G. 2009, \mnras, 400, 398
\bibitem[Kostov et al.(2014)]{Kostov14} Kostov, V. B., McCullough, P. R., Carter, J. A., et al. \ 2014, ApJ, 784, 14 
\bibitem[Leung \& Lee(2013)]{LL13} Leung, G. C. K. \& Lee, M. H. \ 2013, \apj, 763, 107
\bibitem[Lithwick et al.(2012)]{lithwick12} Lithwick, Y., Xie, J., \& Wu, Y. \ 2012, \apj, 761, 122
\bibitem[Liu et al., (2013)]{Liu13} Liu, H.-G., Zhang H. \& Zhou J.-L. \ 2013, ApJL,  767, L38
\bibitem[Mandel \& Agol (2002)]{Mandel02} Mandel, K. \& Agol, E. \ 2002, \apj, 580, 171
\bibitem[{Orosz et al.,(2012a)}]{Orosz12a}Orosz, J. A., Welsh, W. F., Carter, J. A., et al. 2012a, Sci, 337, 1511
\bibitem[{Orosz et al.,(2012b)}]{Orosz12b}Orosz, J. A., Welsh, W. F., Carter, J. A., et al. 2012b, \apj, 758, 87
\bibitem[Ragozzine \& Wolf, (2009)]{Rag09} Ragozzine, D. \& Wolf, A. S., 2009, \apj, 698, 1778	
\bibitem[Schwamb et al.(2013)]{Schwamb13} Schwamb, M. E., Orosz, J. A., Carter, J. A., et al. \ 2013, \apj, 768, 127
\bibitem[Seager \& Hui, (2002)]{Seager02}Seager, S. \& Hui, L. 2002, \apj, 574, 1004
\bibitem[Simon \& Szab{\'o} (2009)]{Simon09} Simon, A. E., Szab{\'o}, Gy. M. \& Szatm{\'a}ry, K. 2009, EM\&P, 105, 385
\bibitem[Steffen et al.(2012a)]{ttvs3} Steffen, J.~H., Fabrycky, D.~C., Ford, E.~B., et al.\ 2012a, \mnras, 421, 2342
\bibitem[Steffen et al.(2012b)]{ttvs6} Steffen, J.~H., Ford, E.~B., Rowe, J.~F., et al.\ 2012b, \apj, 756, 186
\bibitem[Steffen et al.(2013)]{ttvs7} Steffen, J.~H., Fabrycky, D.~C., Agol, E., et al.\ 2013, \mnras, 428, 1077
\bibitem[{Welsh et al.,(2012)}]{Welsh12}Welsh, W. F., Orosz, J. A., Carter, J. A.,  et al. 2012, Nature, 481, 475
\bibitem[Winn et al., (2009)]{Winn09} Winn, J. N., Howard, A. W., Johnson, J. A., et al. 2009, \apj, 703, 2091 
\bibitem[Xie(2012)]{xie12} Xie, J.-W.\ 2012, arXiv:1208.3312
\bibitem[Yang(2013)]{Yang13} Yang, M., Liu, H.-G., Zhang, H., Xie, J.-W. \& Zhou, J.-L., 2013, \apj, 778, 110;
\end{thebibliography}
\end{document}